\documentclass[conference]{IEEEtran}
\usepackage{amsfonts}

\usepackage{amsthm}
\usepackage[fleqn]{amsmath}

\ifCLASSINFOpdf
   \usepackage[pdftex]{graphicx}
\else
   \usepackage[dvips]{graphicx}
   \graphicspath{{../eps/}}
\fi
\begin{document}
%
\title{Exploring Relay Cooperation Scheme for Load-Balance Control in Two-hop Secure Communication System}



%


\author{\IEEEauthorblockN{Yulong Shen\IEEEauthorrefmark{1}\IEEEauthorrefmark{4},
Xiaohong Jiang\IEEEauthorrefmark{2} and Jianfeng
Ma\IEEEauthorrefmark{1}}
\IEEEauthorblockA{\IEEEauthorrefmark{1}School of Computer Science
and Technology, Xidian University, China}
\IEEEauthorblockA{\IEEEauthorrefmark{2}School of Systems Information
Science, Future University Hakodate, Japan}
\IEEEauthorblockA{\IEEEauthorrefmark{4}Email:ylshen@mail.xidian.edu.cn
} }

\maketitle

\begin{abstract}

This work considers load-balance control among the relays under the
secure transmission protocol via relay cooperation in two-hop
wireless networks without the information of both eavesdropper
channels and locations. The available two-hop secure transmission
protocols in physical layer secrecy framework cannot provide a
flexible load-balance control, which may significantly limit their
application scopes. This paper proposes a secure transmission
protocol in case that the path-loss is identical between all pairs
of nodes, in which the relay is randomly selected from the first $k$
preferable assistant relays. This protocol enables load-balance
among relays to be flexibly controlled by a proper setting of the
parameter $k$, and covers the available works as special cases, like
ones with the optimal relay selection ($k=1$) and ones with the
random relay selection ($k = n$, i.e. the number of system nodes).
The theoretic analysis is further provided to determine the maximum
number of eavesdroppers one network can tolerate by applying the
proposed protocol to ensure a desired performance in terms of the
secrecy outage probability and transmission outage probability.

\end{abstract}


%
\IEEEpeerreviewmaketitle

\section{Introduction}

The promising applications of wireless ad hoc networks in many
important scenarios (like battlefield networks, emergency networks,
disaster recovery networks) make a lot of attention turn to ensure
security and high efficiency of wireless transmission. Two-hop
wireless networks, as a building block for large multi-hop network
system, have been a class of basic and important networking
scenarios \cite{IEEEhowto:Sathya}. The analysis and design of secure
transmission protocol in basic two-hop relay networks serves as the
foundation for secure information exchange of general multi-hop
network system. The secure transmission protocols based on
traditional cryptographic approach can hardly achieve everlasting
secrecy, because the adversary can record the transmitted messages
and try any way to break them \cite{IEEEhowto:Talbot}. This
motivates the signaling scheme is considered in physical layer
secrecy framework to provide a strong form of security recently
\cite{IEEEhowto:Wyner}\cite{IEEEhowto:Vasudevan}\cite{IEEEhowto:Koyluoglu}.

By now, a lot of research efforts have been dedicated to secure
transmission through physical layer methods. A few secure
transmission protocols with optimal relay selection via cooperative
relays have been proposed in
\cite{IEEEhowto:Goeckel1}\cite{IEEEhowto:Goeckel2}\cite{IEEEhowto:Vasudevan2},
in which the system node with best link condition to both source and
destination is selected as information relay. Although these
protocols are attractive in the sense that provides very effective
resistance against eavesdroppers, some relay nodes with good link
conditions always preferred to relay package, since the channel
state is relatively constant during a fixed time period, which
results in a severe load-balance problem and a quick node energy
depletion. In fact, the load-balance capacity is of important
property of wireless networks
\cite{IEEEhowto:Hsiao}\cite{IEEEhowto:Gao}. Such, these protocol is
not suitable for energy-limited wireless networks (like wireless
sensor networks). In order to address load-balance problem, Y. Shen
et al. further proposed a random relay selection protocol
\cite{IEEEhowto:Shen1}\cite{IEEEhowto:Shen2}, in which the relay
node is random selected from the system nodes. However, this
protocol has lower transmission efficiency, although achieving a
very good load-balance and energy consumption distribution among
system nodes. Such it is more suitable for large scale wireless
network environment with stringent energy consumption constraint.

In summary, the available secure transmission protocols cannot
provide a flexible load-balance control, which may significantly
limit their application scopes. This paper focuses on this problem
and consider load-balance control capacity for secure transmission
protocol via relay cooperation in two-hop wireless networks without
the information of both eavesdropper channels and locations. The
main contributions of this paper as follows:

\begin{itemize}

\item
This paper extends available works and proposes a secure
transmission protocol via cooperative relays in two-hop relay
wireless networks without the knowledge of eavesdropper channels and
locations, where the relay is randomly selected from the first $k$
preferable assistant relays. With respect to the available works,
this protocol provides flexible load-balance control by a proper
setting of $k$ under the premise of specified secure and reliable
requirements.

\item

The theoretic analysis of the proposed protocol is provided to
determine the corresponding exact results on the maximum number of
eavesdroppers one network can tolerate to satisfy a specified
requirements in case that the path-loss is identical between all
pairs of nodes. The analysis results also show that the proposed
protocol covers all the available secure transmission protocols as
special cases, like ones with the optimal relay selection ($k=1$)
\cite{IEEEhowto:Goeckel1}\cite{IEEEhowto:Goeckel2}\cite{IEEEhowto:Vasudevan2}
and ones with the random relay selection ($k = n$ i.e. the number of
system nodes)\cite{IEEEhowto:Shen1}\cite{IEEEhowto:Shen2}.

\end{itemize}

The remainder of this paper is organized as follows. Section II
presents system models and proposes a secure transmission protocol
with considering load-balance control among the relays. Section III
presents the theoretic analysis of the proposed protocol, and
Section IV concludes this paper.

\section{System Models and Transmission Protocol}

\subsection{Network Model}

A two-hop wireless network scenario is considered where a source
node $S$ wishes to communicate securely with its destination node
$D$ with the help of multiple relay nodes $R_1$, $R_2$, $\cdots$,
$R_n$. Also present in the environment are $m$ eavesdroppers $E_1$,
$E_2$, $\cdots$, $E_m$ without knowledge of channels and locations.
The relay nodes and eavesdroppers are independent and also uniformly
distributed in the network, as illustrated in Fig.1. Our goal here
is to design a transmission protocol to ensure the secure and
reliable information transmission from source $S$ to destination $D$
and provide flexible load-balance control among the relays.

\begin{figure}[!t]
\centering
\includegraphics[width=2.5in]{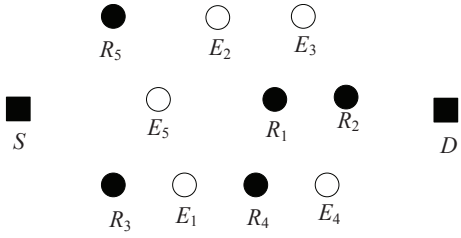}
\DeclareGraphicsExtensions. \caption{System scenario: Source $S$
wishes to communicate securely with destination $D$ with the
assistance of finite relays $R_1$, $R_2$, $\cdots$, $R_{n}$ ($n$=5
in the figure) in the presence of passive eavesdroppers $E_1$,
$E_2$, $\cdots$, $E_{m}$ ($m$=5 in the figure). Cooperative relay
scheme is used in the two-hop transmission.} \label{System scenario}
\end{figure}

\subsection{Transmission Model}

Consider the transmission from a transmitter $A$ to a receiver $B$,
and denote the $i^{th}$ symbol transmitted by node $A$ by
$x_i^{\left(A\right)}$. We assume that all nodes transmit with the
same power $E_s$ and path-loss between all pairs of nodes is
identical and independent. We denote the frequency-nonselective
multi-path fading from $A$ to $B$ by $h_{A,B}$. Under the condition
that all nodes in a group of nodes, $\mathcal {R}$, are generating
noises, the $i^{th}$ signal received at node $B$ from node $A$,
denoted by $y_i^{\left(B\right)}$, is determined as:

$$y_i^{\left(B\right)}=h_{A,B} \sqrt{E_s}x_i^{\left(A\right)} +
\sum_{A_j \in \mathcal {R}}
h_{A_j,B}\sqrt{E_s}x_i^{\left(A_j\right)} + n_i^{\left(B\right)},$$

The multi-path fading $h_{A,B}$ is assumed to follow a Rayleigh
distribution, which remains constant during the transmission of each
packet. Then, $\left|h_{A,B}\right|^2$ is exponentially distributed,
and without loss of generality, we assume that
$E{\left[\left|h_{A,B}\right|^2\right]}=1$. The noise
$n_i^{\left(B\right)}$ at receiver $B$ is assumed to be i.i.d
complex Gaussian random variables with mean $N_0$. The SINR
$C_{A,B}$ from $A$ to $B$ is then given by

$$C_{A,B}=\frac{E_s\left|h_{A,B}\right|^2}{\sum_{A_j \in \mathcal
{R}}E_s{\left|h_{A_j,B}\right|^2}+N_0/2}$$

For a legitimate node and an eavesdropper, we use two separate SINR
thresholds $\gamma_R$ and $\gamma_E$ to define the minimum SINR
required to recover the transmitted messages for legitimate nodes
and eavesdroppers, respectively. Therefore, a system node (the
selected relay or destination) is able to decode a packet if and
only if its received SINR is greater than $\gamma_R$, whereas each
eavesdropper try to achieve target SINR $\gamma_E$ to recover the
transmitted message.

\subsection{Transmission Protocol}

With respect to the protocols with optimal and random relay
selection
\cite{IEEEhowto:Goeckel1}\cite{IEEEhowto:Goeckel2}\cite{IEEEhowto:Vasudevan2}\cite{IEEEhowto:Shen1}
as special cases, a secure transmission protocol is proposed to
enable the tradeoff between the transmission efficiency and
load-balance capacity among the relays to be flexibly controlled.
The proposed protocol works as follows.

\begin{enumerate}

\item
\textbf{\emph{Channel measurement}:} The source $S$ and destination
$D$ broadcast a pilot signal to allow each relay to measure the
channels from $S$ and $D$ to itself. The relays, which receive the
pilot signal, can accurately calculate $h_{S,R_j}$ and $h_{D,R_j},
j=1,2,\cdots,n$.

\item
\textbf{\emph{Candidate relay selection}:} The relays with the first
$k$ large $min\left(|h_{S,R_j}|^2, |h_{D,R_j}|^2\right),
j=1,2,\cdots,n$ form the candidate relay set.

\item
\textbf{\emph{Relay selection}:} The relay, indexed by $j^\ast$, is
selected randomly from candidate relay set. Using the same method
with Step 1, each of the other relays $R_j, j=1,2,\cdots,n, j \neq
j^\ast$ in network exactly knows $h_{R_j,R_{j^\ast}}$.

\item
\textbf{\emph{Two-Hop transmission}:} The source $S$ transmits the
messages to $R_{j^\ast}$, and concurrently, the relay nodes with
indexes in $\mathcal {R}_1 = {\left\{j \neq j^\ast :
|h_{R_j,R_{j^\ast}}|^2 < \tau \right\}}$ transmit noise to generate
interference at eavesdroppers. The relay $R_{j^\ast}$ then transmits
the messages to destination $D$, and concurrently, the relay nodes
with indexes in $\mathcal {R}_2 = {\left\{j \neq j^\ast :
|h_{R_j,D}|^2 < \tau \right\}}$ transmit noise to generate
interference at eavesdroppers.

\end{enumerate}

\emph{Remark 1}: The proposed protocol can flexibly control
load-balance capacity among the relays in terms of networks
requirements by a proper setting of candidate relay set size $k$.
The larger $k$, more system nodes in candidate relay set, means the
better load-balance capacity among the relays and the lower
transmission efficiency, and vice versa.

\emph{Remark 2}: The parameter $\tau$ involved in the proposed
protocol serves as the threshold on path-loss, based on which the
set of noise generating relay nodes can be identified. Notice that a
too large $\tau$ may disable legitimate transmission, while a too
small $\tau$ may not be sufficient for interrupting all
eavesdroppers. Thus, the parameter $\tau$ should be set properly to
ensure both secrecy requirement and reliability requirement.

\section{Theoretical Analysis}

The transmission outage and secrecy outage are first defined to
depict transmission reliability and secrecy, and theoretical
analysis on the maximum numbers of eavesdroppers one network can
tolerate is presented by applying the proposed protocol.

\subsection{Transmission Outage and Secrecy Outage}

For a Two-hop relay transmission from the source $S$ to destination
$D$, we call transmission outage happens if $D$ can not receive the
transmitted packet. We define the transmission outage probability,
denoted by $P_{out}^{\left(T\right)}$, as the probability that
transmission outage from $S$ to $D$ happens. For a predefined upper
bound $\varepsilon_t$ on $P_{out}^{\left(T\right)}$, we call the
communication between $S$ and $D$ is reliable if
$P_{out}^{\left(T\right)} \leq \varepsilon_t$.

Regarding the secrecy outage, we call secrecy outage happens for a
transmission from $S$ to $D$ if at least one eavesdropper can
recover the transmitted packets during the process of this two-hop
transmission. We define the secrecy outage probability, denoted by
$P_{out}^{\left(S\right)}$, as the probability that secrecy outage
happens during the transmission from $S$ to $D$. For a predefined
upper bound $\varepsilon_s$ on $P_{out}^{\left(S\right)}$, we call
the communication between $S$ and $D$ is secure if
$P_{out}^{\left(S\right)} \leq \varepsilon_s$.

\subsection{Analysis of the Proposed Protocol}

We now analyze that under the proposed protocol the number of
eavesdroppers one network can tolerate to ensure the desired
performance in terms of transmission outage probability and secrecy
outage probability. The following two lemmas regarding some basic
properties of $P_{out}^{\left(T\right)}$, $P_{out}^{\left(S\right)}$
and $\tau$ are first presented, which will help us to derive the
main result in Theorem 1.

\emph{Lemma 1}: Consider the network scenario of Fig 1 with equal
path-loss between all pairs of nodes, under the proposed protocol
the transmission outage probability $P_{out}^{\left(T\right)}$ and
secrecy outage probability $P_{out}^{\left(S\right)}$ there satisfy
the following conditions.

\begin{align}
&P_{out}^{\left(T\right)} \leq 2 \left(\frac{1}{k}\sum
\limits_{j=1}^k \bigg[\sum \limits_{i=n-j+1}^n \binom{n}{i}
\left[1-\Psi\right]^{i}\Psi^{n-i}\bigg]\right)\notag\\
&\ \ \ \ -\left(\frac{1}{k}\sum \limits_{j=1}^k \bigg[\sum
\limits_{i=n-j+1}^n \binom{n}{i}
\left[1-\Psi\right]^{i}\Psi^{n-i}\bigg]\right)^2
\end{align}

here, $\Psi =
e^{-2\gamma_R{\left(n-1\right)\left(1-e^{-\tau}\right)}\tau}$ and

\begin{align}
&P_{out}^{\left(S\right)} \leq 2m \cdot \left(\frac{1}{1+\gamma_E}\right)^{\left(n-1\right)\left(1-e^{-\tau}\right)}\notag\\
&\ \ \ \ \ \ \ \ \ -\left[m \cdot
\left(\frac{1}{1+\gamma_E}\right)^{\left(n-1\right)\left(1-e^{-\tau}\right)}\right]^2
\end{align}

Due to space limitation, the proof of this lemma is omitted, which
can be found in the reference \cite{IEEEhowto:Shen3}.

\emph{Lemma 2}: Consider the network scenario of Fig 1 with equal
path-loss between all pairs of nodes, to ensure
$P_{out}^{\left(T\right)} \leq \varepsilon_t$ and
$P_{out}^{\left(S\right)} \leq \varepsilon_s$ under the proposed
protocol, the parameter $\tau$ must satisfy the following condition.

\begin{align*}
\tau \leq \sqrt{\frac{ -\log\left(
\left[\binom{k}{\lfloor\frac{k}{2}\rfloor}\left(1 +
k\sqrt{1-\varepsilon_t}\right)\right]^{\frac{1}{k}} - 1
\right)}{2\gamma_R\left(n-1\right)}}
\end{align*}
and

\begin{align*}
\tau \geq - \log{\left[1 + \frac{\log{\left(\frac{1 - \sqrt{1 -
\varepsilon_s}}{m}\right)}}{\left(n - 1\right)\log{\left(1 +
\gamma_E\right)}}\right]}
\end{align*}

\begin{proof}

The parameter $\tau$ should be set properly to satisfy both
reliability and secrecy requirements.

\textbf{$\bullet$ Reliability Guarantee}

To ensure the reliability requirement $P_{out}^{\left(T\right)} \leq
\varepsilon_t$, we know from formula (1) in the Lemma 1, that we
just need

\begin{align*}
&2 \left(\frac{1}{k}\sum \limits_{j=1}^k \bigg[\sum
\limits_{i=n-j+1}^n \binom{n}{i}
\left[1-\Psi\right]^{i}\Psi^{n-i}\bigg]\right)\\
&-\left(\frac{1}{k}\sum \limits_{j=1}^k \bigg[\sum
\limits_{i=n-j+1}^n \binom{n}{i}
\left[1-\Psi\right]^{i}\Psi^{n-i}\bigg]\right)^2\\
&\leq \varepsilon_t
\end{align*}

Thus,

\begin{equation}
\begin{aligned}
&\frac{1}{k}\sum \limits_{j=1}^k \bigg[\sum \limits_{i=n-j+1}^n
\binom{n}{i} \left[1-\Psi\right]^{i}\Psi^{n-i}\bigg] \leq 1-
\sqrt{1-\varepsilon_t}
\end{aligned}
\end{equation}

Notice that

\begin{equation}
\begin{aligned}
&\frac{1}{k}\sum \limits_{j=1}^k \bigg[\sum \limits_{i=n-j+1}^n
\binom{n}{i}\left(1-\Psi\right)^{i}\Psi^{n-i}\bigg]\\
& = \frac{1}{k}\sum \limits_{j=1}^k \bigg[1 - \sum
\limits_{i=0}^{n-j}
\binom{n}{i}\left(1-\Psi\right)^{i}\Psi^{n-i}\bigg]\\
& = \frac{1}{k}\sum \limits_{j=1}^k \bigg[1 - \sum
\limits_{i=0}^{n-j}
\frac{\binom{n}{i}}{\binom{n-j}{i}}\binom{n-j}{i}\left(1-\Psi\right)^{i}\Psi^{n-j-i
}\Psi^j\bigg]
\end{aligned}
\end{equation}

We also notice $i$ can take from $0$ to $n-j$, then we have

\begin{equation*}
\begin{aligned}
& 1 \leq \frac{\binom{n}{i}}{\binom{n-j}{i}} \leq \frac{n!}{(n-j)!j!}\\
\end{aligned}
\end{equation*}

Substituting into formula (4), we have

\begin{equation}
\begin{aligned}
&\frac{1}{k}\sum \limits_{j=1}^k \bigg[1 - \sum \limits_{i=0}^{n-j}
\frac{\binom{n}{i}}{\binom{n-j}{i}}\binom{n-j}{i}\left(1-\Psi\right)^{i}\Psi^{n-j-i
}\Psi^j\bigg]\\
& \leq \frac{1}{k}\sum \limits_{j=1}^k \bigg[1 - \Psi^j \cdot \sum
\limits_{i=0}^{n-j}\binom{n-j}{i}\left(1-\Psi\right)^{i}\Psi^{n-j-i}\bigg]\\
&= 1 - \frac{1}{k}\sum \limits_{j=1}^k  \Psi^j \\
&= 1 - \frac{1}{k}\bigg[\sum \limits_{j=0}^k \frac{1}{\binom{k}{j}} \binom{k}{j} \Psi^j - 1\bigg]\\
&\leq 1 - \frac{1}{k}\bigg[\frac{1}{\binom{k}{\lfloor\frac{k}{2}\rfloor}} \sum \limits_{j=0}^k  \binom{k}{j} \Psi^j - 1\bigg]\\
&= 1 - \frac{1}{k}\bigg[\frac{1}{\binom{k}{\lfloor\frac{k}{2}\rfloor}} (1+\Psi)^k - 1\bigg]\\
\end{aligned}
\end{equation}

According to formula (3), (4) and (5), in order to ensure the
reliability, we need

\begin{equation*}
1 - \frac{1}{k}\bigg[\frac{1}{\binom{k}{\lfloor\frac{k}{2}\rfloor}}
(1+\Psi)^k - 1\bigg] \leq 1- \sqrt{1-\varepsilon_t}
\end{equation*}

or equally,

\begin{equation*}
\Psi \geq \left[\binom{k}{\lfloor\frac{k}{2}\rfloor}\left(1 +
k\sqrt{1-\varepsilon_t}\right)\right]^{\frac{1}{k}} - 1
\end{equation*}

that is,

\begin{equation*}
e^{-2\gamma_R\left(n-1\right) \cdot \left(1-e^{-\tau}\right)\tau}
\geq \left[\binom{k}{\lfloor\frac{k}{2}\rfloor}\left(1 +
k\sqrt{1-\varepsilon_t}\right)\right]^{\frac{1}{k}} - 1
\end{equation*}

Therefore

\begin{equation*}
\left(1-e^{-\tau}\right)\tau \leq \frac{ -\log\left(
\left[\binom{k}{\lfloor\frac{k}{2}\rfloor}\left(1 +
k\sqrt{1-\varepsilon_t}\right)\right]^{\frac{1}{k}} - 1
\right)}{2\gamma_R\left(n-1\right) }
\end{equation*}

By using Taylor formula, we have

\begin{align*}
\tau \leq \sqrt{\frac{ -\log\left(
\left[\binom{k}{\lfloor\frac{k}{2}\rfloor}\left(1 +
k\sqrt{1-\varepsilon_t}\right)\right]^{\frac{1}{k}} - 1
\right)}{2\gamma_R\left(n-1\right)}}
\end{align*}

\textbf{$\bullet$ Secrecy Guarantee}

To ensure the secrecy requirement $P_{out}^{\left(S\right)} \leq
\varepsilon_s$, we know from formula (2) in Lemma 1, that we just
need

\begin{align*}
&2m \cdot \left(\frac{1}{1+\gamma_E}\right)^{\left(n-1\right)\left(1-e^{-\tau}\right)}\\
& -\left[m \cdot
\left(\frac{1}{1+\gamma_E}\right)^{\left(n-1\right)\left(1-e^{-\tau}\right)}\right]^2\\
& \leq \varepsilon_s
\end{align*}

Thus,

\begin{align*}
&m \cdot
\left(\frac{1}{1+\gamma_E}\right)^{\left(n-1\right)\left(1-e^{-\tau}\right)}
\leq 1- \sqrt{1-\varepsilon_s}
\end{align*}

That is,

\begin{align*}
& \tau \geq - \log{\left[1 + \frac{\log{\left(\frac{1 - \sqrt{1 -
\varepsilon_s}}{m}\right)}}{\left(n - 1\right)\log{\left(1 +
\gamma_E\right)}}\right]}
\end{align*}

\end{proof}

Based on the results of Lemma 2, we now can establish the following
theorem regarding the performance of the proposed protocol in case
of equal path-loss between all node pairs.

\textbf{Theorem 1.}  Consider the network scenario of Fig 1 with
equal path-loss between all pairs of nodes. To guarantee
$P_{out}^{\left(T\right)} \leq \varepsilon_t$ and
$P_{out}^{\left(S\right)} \leq \varepsilon_s$ under the proposed
protocol, the number of eavesdroppers $m$ the network can tolerate
must satisfy the following condition.

\begin{align*}
& m \leq \frac{1 - \sqrt{1 -
\varepsilon_s}}{\left(\frac{1}{1+\gamma_E}\right)^{\sqrt{\frac{-\left(n-1\right)\log\left(
\left[\binom{k}{\lfloor\frac{k}{2}\rfloor}\left(1 +
k\sqrt{1-\varepsilon_t}\right)\right]^{\frac{1}{k}} - 1
\right)}{2\gamma_R}}}}
\end{align*}

$\lfloor \cdot \rfloor$ is the floor function.

\begin{proof}

From Lemma 2, we know that to ensure the reliability requirement, we
have

\begin{align}
\tau \leq \sqrt{\frac{ -\log\left(
\left[\binom{k}{\lfloor\frac{k}{2}\rfloor}\left(1 +
k\sqrt{1-\varepsilon_t}\right)\right]^{\frac{1}{k}} - 1
\right)}{2\gamma_R\left(n-1\right)}}
\end{align}

and

\begin{equation}
\left(n-1\right)\left(1-e^{-\tau}\right) \leq \frac{ -\log\left(
\left[\binom{k}{\lfloor\frac{k}{2}\rfloor}\left(1 +
k\sqrt{1-\varepsilon_t}\right)\right]^{\frac{1}{k}} - 1
\right)}{2\gamma_R\tau}
\end{equation}

To ensure the secrecy requirement, we need

\begin{align}
& \left(\frac{1}{1+\gamma_E}\right)^{\left(n
-1\right)\left(1-e^{-\tau}\right)} \leq \frac{1 - \sqrt{1 -
\varepsilon_s}}{m}
\end{align}

From formula (7) and (8), we can get

\begin{align}
& m \leq \frac{1 - \sqrt{1 -
\varepsilon_s}}{\left(\frac{1}{1+\gamma_E}\right)^{\left(n
-1\right)\left(1-e^{-\tau}\right)}}\notag\\
&\ \ \ \ \leq \frac{1 - \sqrt{1 -
\varepsilon_s}}{\left(\frac{1}{1+\gamma_E}\right)^{\frac{
-\log\left( \left[\binom{k}{\lfloor\frac{k}{2}\rfloor}\left(1 +
k\sqrt{1-\varepsilon_t}\right)\right]^{\frac{1}{k}} - 1
\right)}{2\gamma_R\tau}}}
\end{align}

By letting $\tau$ take its maximum value for maximum interference at
eavesdroppers, from formula (6) and (9), we get the following bound

\begin{align*}
& m \leq \frac{1 - \sqrt{1 -
\varepsilon_s}}{\left(\frac{1}{1+\gamma_E}\right)^{\sqrt{\frac{-\left(n-1\right)\log\left(
\left[\binom{k}{\lfloor\frac{k}{2}\rfloor}\left(1 +
k\sqrt{1-\varepsilon_t}\right)\right]^{\frac{1}{k}} - 1
\right)}{2\gamma_R}}}}
\end{align*}

\end{proof}

Based on the above analysis, by simple derivation, we can get the
follow corollary to show our proposal is a general protocol.

\textbf{Corollary 1.} Consider the network scenario of Fig 1 with
equal path-loss between all pairs of nodes, the analysis results of
the proposed protocol is identical to that of protocols with the
optimal relay selection presented in
\cite{IEEEhowto:Goeckel1}\cite{IEEEhowto:Goeckel2} by setting of $k
= 1$, and is identical to that of protocols with the random relay
selection presented in \cite{IEEEhowto:Shen1}\cite{IEEEhowto:Shen2}
by setting of $k = n$. More general, the proposed protocol can
flexibly control load-balance capacity among the relays in terms of
networks requirements by a proper setting of candidate relay set
size $k$ in case that the path-loss is equal between all pairs of
nodes.

\section{Conclusion}

This paper extended the available protocol with considering
load-balance capacity and proposed a general protocol to ensure
secure and reliable information transmission through multiple
cooperative system nodes for two-hop relay wireless networks without
the knowledge of eavesdropper channels and locations. We proved that
the proposed protocol has the ability of flexible load-balance
control by a proper setting of the size $k$ of candidate relay set.
Such, in general it is possible for us to set proper value of
parameters according to network scenario to support various wireless
applications.







%

\end{document}